\documentstyle{article}

\newcommand{\beq}{\begin{equation}}
\newcommand{\eeq}{\end{equation}}
\newcommand{\ba}{\begin{array}}
\newcommand{\ea}{\end{array}}
\newcommand{\bea}{\begin{eqnarray}}
\newcommand{\eea}{\end{eqnarray}}
\newcommand{\bean}{\begin{eqnarray*}}
\newcommand{\eean}{\end{eqnarray*}}

\begin{document}

\begin{center}

{\Large \sc \bf INITIAL-BOUNDARY VALUE PROBLEMS}

\vskip 5pt

{\Large \sc \bf FOR LINEAR AND SOLITON  PDEs} 

\vskip 20pt

{\large  A. Degasperis$^{1,2,\S}$, S. V. Manakov$^{3,\S}$ and 
P. M. Santini$^{1,2,\S}$}

\bigskip

{\it
 $^1$Dipartimento di Fisica, Universit\`a di Roma "La Sapienza"\\
Piazz.le Aldo Moro 2, I-00185 Roma, Italy

\smallskip

$^2$Istituto Nazionale di Fisica Nucleare, Sezione di Roma\\
P.le Aldo Moro 2, I-00185 Roma, Italy

\smallskip

$^3$Landau Institute for Theoretical Physics, Moscow, Russia}

\bigskip

$^{\S}$e-mail: ~{\tt antonio.degasperis@roma1.infn.it,~ manakov@itp.ac.ru, \\ 
paolo.santini@roma1.infn.it}

\bigskip

\end{center}

\begin{abstract}
Evolution PDEs for dispersive waves are considered in both linear and 
nonlinear integrable cases, and initial-boundary value problems 
associated with them are formulated in spectral space. A method of solution 
is presented, which is based on the elimination of the unknown boundary 
values by proper restrictions of the functional space and of the 
spectral variable complex domain. 
Illustrative examples include the linear 
Schr\"odinger equation on compact and semicompact 
$n$-dimensional domains and the nonlinear Schr\"odinger equation on the 
semiline.
\end{abstract}

\vskip 20pt

\section{Introduction}

Initial-Boundary Value (IBV) problems for Partial Differential Equations 
(PDEs) play an important role in applications to physics and, in general, 
to natural sciences. 

It is well known that the basic difficulty associated with the study of 
IBV problems for linear and nonlinear integrable PDEs is the presence 
of unknown 
boundary values in the relevant equations of any method of solution. 
In this paper, after formulating the IBV problems for linear and nonlinear 
integrable PDEs in spectral space, we present a method of solution, the 
Elimination-by-Restriction (EbR) approach, which is based 
on a strategy of elimination of the unknown boundary values by proper 
restriction of the functional space and of the 
complex domain of definition of the involved spectral 
functions. This approach is inspired by the Green's function (GF) method, 
which, in the linear context, is essentially its counterpart in 
configuration space, and by our recent findings in the nonlinear context. 
The paper is organized as follows. In \S 2 we deal with IBV problems for 
linear PDEs with constant coefficients. After introducing the proper 
Fourier Transform (FT) for that problem and establishing 
its analyticity properties, we first express the Fourier transform of 
the solution in terms of the 
Fourier transforms of known and unknown initial - boundary 
values using Green's formula. Then we present the EbR approach in which, 
using systematically a strategy of elimination of the unknown boundary 
values,  
one obtains the appropriate spectral representation of the solution 
whose support may eventually turn out to be discrete rather than 
continuous as in the general Fourier integral one starts from. We 
illustrate the power of the method solving IBV problems for the 
Schr\"odinger equation on an $n$-dimensional rectangular box and quadrant. 
In \S 3 we apply the EbR strategy to soliton equations. After 
defining the proper spectral transform $S(k,t)$ for the given IBV problem, 
we apply the EbR procedure eliminating the unknown boundary values from the 
equations defining $S(k,t)$ and characterizing $S(k,t)$ via a nonlinear 
integral equation. This approach, presented on the prototype example 
of the nonlinear Schr\"odinger (NLS) equation on the semiline, can be 
in principle generalized to the segment case. Since the EbR technique 
operates on the spectral variables conjugated to the space ones, it 
works well either using the space-time transform or just the space transform. 
In the linear case we find it simpler to work with the space-time FT, 
while in the nonlinear case it seems more convenient to use the space 
transform, namely the well-known Inverse Scattering (Spectral) Transform 
(IST). We finally show the equivalence between IBV problems for soliton 
equations on the semiline and some specific  
forced initial value problems on the whole line. An important 
application of this equivalence is that, from the well-known asymptotics of 
soliton equations on the whole line, one can obtain immediately 
the asymptotic behaviour for 
IBV problems on the semiline with decaying boundary data. We have confined 
the relevant literature to \S 4.

The results contained in this paper are an expansion of part of    
the  material presented by the authors at the Euroconference 
``NEEDS 2001'' and at the 
workshop:  ``Boundary Value Problems'', the first and last events of the 
Semester: {\it Integrable Systems}, held at the Isaac Newton Institute 
of Cambridge during the period July - December 2001.

\section{\bf The Elimination-by-Restriction Approach: \break The Linear Case}

It is well -known that the Fourier Transform (FT) is the proper tool to solve 
initial - boundary value (IBV) problems for linear PDE's in ${\cal R}^{n+1}$ 
with decaying boundary values:
\bea\label{eq:IBV1}
\ba{c}
{\cal L}({\bf \bigtriangledown},{\partial\over\partial t})u({\bf x},t)=
f({\bf x},t),~~~~{\bf x}=(x_1,..,x_n)\in {\cal R}^{n},~~~t>0,  \\
u({\bf x},0)=u_0({\bf x}),~~~~u({\bf x},t)\to 0,~|{\bf x}|\to\infty ,
\ea
\eea
where ${\bf \bigtriangledown} =({\partial\over\partial {x_1}},\cdot\cdot ,
{\partial\over\partial {x_n}})$, ${\cal L}$ is a constant coefficients 
partial differential operator, $u({\bf x},t)$ is the unknown field, 
$f({\bf x},t)$ is a given forcing and $u_0({\bf x})$ is the given initial 
condition.

In this section we present a 
very  effective approach, in Fourier space,  
for solving  
more complicated IBV problems, defined on compact or semi - compact 
space domains $V$: 
\beq\label{eq:IBV2}
{\cal L}({\bf \bigtriangledown},{\partial\over\partial t})u({\bf x},t)=
f({\bf x},t),
~~~~~{\bf x}\in V\subset {\cal R}^n,~~t>0, 
\eeq
with Dirichelet or Neumann or mixed boundary conditions on $\partial V$.

\subsection{\bf The Fourier Transform and its properties}

The natural FT associated with the space - time domain 
${\cal D}=V\otimes (0,\infty )$ (in short: $FT_{\cal D}$) is defined by
\beq\label{eq:FT}
\hat F({\bf k},q)=\int_{\cal D}d{\bf x}dt
e^{-i({\bf k}\cdot {\bf x}+qt)}F({\bf x},t)
\eeq 
for any smooth function $F({\bf x},t)$, $({\bf x},t)\in {\cal D}$, assuming 
that $F({\bf x},t)\to 0,~t\to\infty$ fast enough; here 
${\bf k}=(k_1,..,k_n)\in {\cal R}^{n},~q\in {\cal R}$ and 
${\bf k}\cdot {\bf x}=\sum_jk_jx_j$. Its inverse:
\beq\label{eq:invFT} 
F({\bf x},t)\chi_{\cal D}({\bf x},t)=
\int_{{\cal R}^{n+1}}{d{\bf k}dq\over (2\pi )^{n+1}}
e^{i({\bf k}\cdot {\bf x}+qt)}\hat F({\bf k},q)
\eeq
reconstructs $F({\bf x},t)$ in $\cal D$ and zero outside, 
where $\chi_{\cal D}({\bf x},t)$ is the characteristic 
function of the domain $\cal D$: $\chi_{\cal D}({\bf x},t)=1,~({\bf x},t)\in 
{\cal D},~\chi_{\cal D}({\bf x},t)=0,~({\bf x},t)\notin {\cal D}$ 
(therefore: $\chi_{\cal D}({\bf x},t)=
\chi_{V}({\bf x})H(t)$, where $H(t)$ is the usual Heaviside (step) 
function). 

If the space domain is the whole space: $V={\cal R}^n$, the $FT_{\cal D}$ 
(\ref{eq:FT}) is defined in ${\cal A}={\cal R}^n\otimes \bar{\cal I}_q$, where 
$\bar{\cal I}_q$ is the closure of the lower half $q$-plane ${\cal I}_q$, 
analytic in $q\in {\cal I}_q,~\forall 
{\bf k}\in {\cal R}^n$ and exhibits a proper asymptotic 
behaviour for large $|q|$ in the analyticity region.  
If the space domain $V$ is compact, the $FT_{\cal D}$ acquires strong 
analyticity properties in all the Fourier variables: it is defined in 
${\cal A}={\cal C}^n\otimes\bar{\cal I}_q$,
 analytic in $q\in {\cal I}_q,~\forall {\bf k}\in {\cal C}^n$, entire  
in every complex $k_j,~j=1,..,n$ $\forall q\in \bar{\cal I}_q$ and  
exhibits a proper asymptotic behaviour, for 
large $({\bf k},q)$, in the analyticity regions.  
If the space domain is semi - compact, then the analyticity in the Fourier 
variables $k_j,~j=1,..,n$ is limited to open regions of the complex plane, 
depending on the geometric properties of the domain $V$.

To express the $FT_{\cal D}$ of the solution in terms of the $FT_{\cal D}$'s 
of the forcing and of 
the IB conditions we make use of the well - known {\bf Green's formula 
(identity)}:

\beq\label{eq:Gformula}
b{\cal L}a-a \tilde {\cal L} b=div ~J({\bf x},t),
\eeq
and of its integral consequence, the celebrated {\bf Green's integral 
identity}:
\beq\label{eq:Gint}
\int_{{\cal D}}(b{\cal L}a-a \tilde {\cal L} b)d{\bf x}dt=
\int_{\partial {\cal D}}J({\bf x},t)\cdot {\bf \nu} d\sigma , 
\eeq
obtained by integrating (\ref{eq:Gformula}) over the domain ${\cal D}$ and 
by using the divergence theorem. In equation (\ref{eq:Gformula}),  
$\tilde {\cal L}$ is the formal adjoint of ${\cal L}$:  
$\tilde {\cal L}= {\cal L}(-{\bf \bigtriangledown},
-{\partial\over\partial t})$, $J({\bf x},t)$ is an $(n+1)$-dimensional 
vector field, 
$div$ is the $(n+1)$-dimensional divergence operator and     
$a({\bf x},t)$ and $b({\bf x},t)$ are arbitrary functions.  In equation 
(\ref{eq:Gint}),   
$d\sigma$ is the hypersurface element of the boundary and $\bf \nu$ is 
its outward unit normal. We remark that, given ${\cal L}$, its 
formal adjoint $\tilde {\cal L}$ and two arbitrary functions $a$ and $b$, 
an $(n+1)$-dimensional vector field $J({\bf x},t)$ satisfying 
the Green's formula (\ref{eq:Gformula}) always exists and can be  
algorithmically found to be a linear expression of $a$, $b$ and their 
partial derivatives of order up to $N-1$, if $\cal L$ is of order $N$. 

The arbitrariness of $a$ and $b$ allows one to extract from 
(\ref{eq:Gformula}) and (\ref{eq:Gint}) several important informations on the 
BV problem; with the particular choice  
\beq\label{eq:choice1}
a=u({\bf x},t),~~~~~~~~~b=e^{-i({\bf k}\cdot{\bf x}+qt)}/{\cal L}(i{\bf k},iq),
\eeq
where ${\cal L}(i{\bf k},iq)$ is the eigenvalue of the operator $\cal L$, 
corresponding to the eigenfunction $e^{i({\bf k}\cdot {\bf x}+qt)}$,   
the vector field $J$ takes the following form: \break
$J=e^{-i({\bf k}\cdot{\bf x}+qt)}J'({\bf x},t;
{\bf k},q)/{\cal L}(i{\bf k},iq)$ and 
the Green's integral identity (\ref{eq:Gint}) gives the $FT_{\cal D}$ 
of the solution in terms of the $FT_{\cal D}$'s (or, maybe, of 
generalized FT's) of the forcing 
and of all the initial - boundary values:
\beq\label{eq:FTu}
\hat u({\bf k},q)={\hat f({\bf k},q)-
\int_{\partial {\cal D}}e^{-i({\bf k}\cdot {\bf x}+qt)}
J'({\bf x},t;{\bf k},q)\cdot \nu d\sigma \over 
{\cal L}(i{\bf k},iq)}=:
{\hat{\cal N}({\bf k},q)\over {\cal L}(i{\bf k},iq)},~~~
({\bf k},q)\in{\cal A}.  
\eeq
In general, ${\cal L}(i{\bf k},iq)$, the denominator of the above equation, is 
an entire and, most frequently, polynomial function of all its complex 
variables and its zeroes may lie on the real axis; therefore, before 
calculating the inverse FT, we must regularize it:
\beq\label{eq:reg}
{\cal L}(i{\bf k},iq)\to {\cal L}_{reg}(i{\bf k},iq);
\eeq
i.e., we must move a bit the singularities off the real axis, outside 
the domain $\cal A$.

Its inverse transform (\ref{eq:invFT}) gives the corresponding {\bf Fourier 
representation} of the solution:
\beq\label{eq:Frepr}
U({\bf x},t)=
u({\bf x},t)\chi_{\cal D}({\bf x},t)=\int_{{\cal R}^{n+1}}{d{\bf k}dq\over 
(2\pi )^{n+1}}e^{i({\bf k}\cdot {\bf x}+qt)}
{\hat{\cal N}({\bf k},q)\over {\cal L}_{reg}(i{\bf k},iq)},~~
({\bf x},t)\in{\cal R}^{n+1}. 
\eeq
Clearly this is not the end of the story since, in general, the RHS of 
equation (\ref{eq:FTu}) depends on known and unknown boundary values. 

\subsection{\bf Elimination-by-Restriction in Fourier space}

The traditional ways in which IBV problems for linear PDE's are solved 
consist in finding convenient strategies for  {\it eliminating the unknown 
boundary conditions} from the representation of the solution. On this idea 
is based the celebrated Green's function (GF) approach, in which: 

\noindent
i) one constructs the Green's integral representation 
\beq\label{eq:Gintrepr}
u({\bf x},t)=\int_{\cal D}d{\bf x}'dt' 
\tilde g({\bf x},t;{\bf x}',t')f({\bf x}',t')-
\int_{\partial{\cal D}}J({\bf x},t;{\bf x}',t')\cdot 
{\nu}_{x'}d\sigma_{x'} ,~~
({\bf x},t)\in {\cal D}. 
\eeq
of the solution of the IBV problem (\ref{eq:IBV2}) as another 
application of (\ref{eq:Gint}), corresponding to the choice: 
$a({\bf x},t)= u({\bf x},t)$ and 
$b({\bf x},t)=\tilde g({\bf x}',t';{\bf x},t)$, where  $\tilde g$ is {\bf any} 
Green's function of $\tilde {\cal L}_x$: $\tilde {\cal L}_x
\tilde g({\bf x}',t';{\bf x},t)=\delta ({\bf x}-{\bf x}')\delta (t-t'),~~
({\bf x},t),~({\bf x}',t')\in {\cal D}$. 
ii) One uses the arbitrariness of $\tilde g$ and constructs that particular 
Green's function which 
allows one to eliminate contributions depending on unknown boundary 
values.  
On the elimination idea is also based the eigenfunction expansion method, 
essentially 
equivalent to the GF approach, in which one constructs a set of 
eigenfunctions of $\cal L$ with proper boundary conditions 
which allow again to eliminate the unknown boundary values. Both approaches 
are of functional analytical nature.

In this section we shall show how the elimination strategy can be 
conveniently implemented    
working in the Fourier space defined by (\ref{eq:FT}) (the elimination 
strategy in spectral space has been already investigated for 
soliton equations on the semiline (see \S 4)). 

Equation (\ref{eq:FTu}), defined in a proper domain 
${\cal A}\subset {\cal C}^{n+1}$, usually exhibits several symmetry 
properties which are consequence of the structure of $\cal L$ and of the 
geometry of the space domain $V$. 
The first and more critical part of the 
method consists in constructing a linear operator $\cal E$ which, 
by exploiting systematically these 
symmetry properties in Fourier space, 
annihilates, in the RHS of (\ref{eq:FTu}), 
the contributions coming from the unknown boundary values:
\beq\label{eq:elim}
{\cal E}\hat u({\bf k},q)={\cal E}({\hat N\over {\cal L}})({\bf k},q) =
\{only~known~quantities\}, ~~~({\bf k},q)\in{\cal A}'\subset{\cal A}.
\eeq    
The elimination procedure, described by the linear operator $\cal E$, allows 
one to construct ${\cal E}\hat u$ in a subspace ${\cal A}'$ of the original 
space of definition of $\hat u$. It is necessary therefore to establish if 
this information is sufficient to reconstruct the solution $u$; i.e., if 
${\cal E}\hat u$ defines an invertible spectral transform in $\cal D$.

Using equation (\ref{eq:FT}), this new spectral transform is defined by:
\bea\label{eq:newT}
\ba{c}
{\cal E}\hat u({\bf k},q)=\int_{\cal D}d{\bf x}dt
\tilde\varphi_{{\bf k},q}({\bf x},t)
u({\bf x},t),~~~~~~~~({\bf k},q)\in{\cal A}'  \\
\tilde\varphi_{{\bf k},q}({\bf x},t):=
{\cal E}(e^{-i({\bf k}\cdot {\bf x}+qt)}), 
\ea
\eea
and  is invertible provided   
an eigenfunction $\varphi_{{\bf k},q}({\bf x},t)$ of $\cal L$ 
exists with the property of satisfying the completeness condition:
\beq\label{eq:compl}
\sum_{({\bf k},q)\in{\cal A}'}\varphi_{{\bf k},q}({\bf x},t)
\tilde\varphi_{{\bf k},q}({\bf x}',t')=\delta (t-t')\delta ({\bf x}-{\bf x}'),
~~~~({\bf x},t),({\bf x}',t')\in {\cal D}.
\eeq
In this case the inverse transform allows one to construct a function 
$\tilde U({\bf x},t)$, defined in general on the whole space-time, 
which coincides with the solution in $\cal D$:
\bea\label{eq:U}
\ba{c}
\tilde U({\bf x},t)=
\sum\limits_{{\bf k},q}\varphi_{{\bf k},q}({\bf x},t)
{\cal E}({\hat N\over {\cal L}})({\bf k},q),~~~
({\bf x},t)\in{\cal R}^{n}\otimes (0,\infty ),  \\
\tilde U({\bf x},t)=u({\bf x},t),~~~~~({\bf x},t)\in{\cal D}.
\ea
\eea
In equations (\ref{eq:compl}) and (\ref{eq:U}) 
$\sum_{{\bf k},q}$ indicates a sum and/or an integral, depending on the 
nature of the domain ${\cal A}'$ of definition of the new 
spectral transform. Summarizing:

\noindent
{\bf Although the direct transform we started with 
is the Fourier transform $\hat u$ defined in (\ref{eq:FT}), 
the elimination-by-restriction procedure performed on it leads to 
a new direct  
transform ${\cal E}\hat u$ (\ref{eq:newT}), whose inversion usually 
differs considerably from (\ref{eq:invFT}). This implies that the 
reconstructed function $\tilde U({\bf x},t)$ coincides with the solution 
$u({\bf x},t)$ in 
$\cal D$ but is not zero outside $\cal D$, inheriting the symmetry 
properties of the eigenfunction $\varphi_{{\bf k},q}({\bf x},t)$}. 

\vskip 5pt
\noindent

\subsection{Illustrative Example} 
In this section we apply the EbR approach to the IBV problem 
(\ref{eq:IBV2}) corresponding to 
\bea\label{eq:Lex}
\ba{c}
{\cal L}=
i{\partial\over \partial t}+\bigtriangleup , ~~~~~~~~~
\bigtriangleup =\bigtriangledown\cdot\bigtriangledown =
\sum\limits_{j=1}^n{\partial^2\over\partial {x_j}^2}, \\
V=\{{\bf x}:~0\le x_j\le L_j,~j=1,..,n\};
\ea
\eea
i.e., we solve IBV problems for the $(n+1)$-dimensional Schr\"odinger 
equation in an $n$-dimensional rectangular box.
 
Then:
\bea\label{eq:Jex}
\ba{c}
\tilde {\cal L}=-i{\partial\over \partial t}+\bigtriangleup ,~~~~~~ 
J=(iab,b\bigtriangledown a-a\bigtriangledown b), \\
{\cal L}(i{\bf k},iq)=-(q+k^2)~~\Rightarrow~~{\cal L}_{reg}(i{\bf k},iq)=
-(q+k^2-i0),
\ea
\eea
where $k^2={\bf k}\cdot {\bf k}$. 
Equations (\ref{eq:FTu}) and (\ref{eq:Jex}) give the following 
expression of the Fourier transform of the solution 
in terms of the Fourier transforms of the 
forcing and of all the initial - boundary values:
\bea\label{eq:FTuboxn}
\ba{c}
\hat u({\bf k},q)=-{\hat{\cal N}({\bf k},q)\over q+k^2-i0},~~~~~~~~~~~~~
({\bf k},q)\in{\cal A},  \\
\hat{\cal N}({\bf k},q):=\hat f({\bf k},q)+i\hat u_0({\bf k})+
\sum\limits_{j=1}^n\{ [\hat w_{0j}({\bf k}_j,q)+
ik_j\hat v_{0j}({\bf k}_j,q)]-  \\
e^{-ik_jL_j}[\hat w_{Lj}({\bf k}_j,q)+ik_j\hat v_{Lj}({\bf k}_j,q)] \},
\ea
\eea
in the definition domain ${\cal A}={\cal C}^n\otimes\bar{\cal I}_q$, 
where $\hat f,~\hat u_0,~\hat v_{0j},~\hat v_{Lj},~\hat w_{0j},
~\hat w_{Lj}$ are the FTs of the forcing and of the following IB values:
\bea\label{eq:IB}
\ba{c}
u_0({\bf x})=u({\bf x},t)|_{t=0},~~
v_{0j}({\bf x}_j,t)=u({\bf x},t)|_{x_j=0},~~
v_{Lj}({\bf x}_j,t)=u({\bf x},t)|_{x_j=L_j},~~\\
w_{0j}({\bf x}_j,t)={\partial u\over \partial x_j}({\bf x},t)|_{x_j=0},~~
w_{Lj}({\bf x}_j,t)={\partial u\over \partial x_j}({\bf x},t)|_{x_j=L_j};
\ea
\eea 
i.e., for instance:
\bea\label{eq:FTIB}
\ba{c}
\hat u_0({\bf k})=\int\limits_Vd{\bf x}e^{-i{\bf k}\cdot{\bf x}}u_0({\bf x}),
~~\hat v_{0j}({\bf k}_j,q)=
\int\limits_0^{\infty}dt\int\limits_{V_j}d{\bf x}_j
e^{-i({\bf k}_j\cdot{\bf x}_j+qt)}v_{0j}({\bf x}_j,t).
\ea
\eea
In equations (\ref{eq:FTuboxn})-(\ref{eq:FTIB}), as well as in the following,  
${\bf x}_j=(x_1,..,\check{x_j},..,x_n)\in {\cal R}^{n-1}$, 
${\bf k}_j=(k_1,..,\check{k_j},..,k_n)\in {\cal R}^{n-1}$,  
$\int_{V_j}d{\bf x}_j=\int_0^{L_1}dx_1\cdot\cdot(\check{\int_0^{L_j}dx_j})
\cdot\cdot\int_0^{L_n}dx_n$ with the understanding that 
the superscript $\check{~}$ indicates that the quantity underneath is 
removed. 
 
Of course, if all the above boundary values were known, the solution $u$ 
would be given by the formula (\ref{eq:Frepr}):

\bea\label{eq:Freprboxn}
\ba{c}
U({\bf x},t)=
u({\bf x},t)\chi_V({\bf x})H(t)=
-\int_{{\cal R}^{n+1}}{d{\bf k}dq\over (2\pi )^{n+1}}
e^{i({\bf k}\cdot {\bf x}+qt)}{\hat{\cal N}({\bf k},q)\over q+k^2-i0}= \\
-\int_{{\cal R}^{n+1}}{dqd{\bf k} \over (2\pi )^{n+1}}
e^{i({\bf k}\cdot {\bf x}+qt)}
{\hat f({\bf k},q)\over q+k^2-i0}
+H(t)\int_{{\cal R}^n}{d{\bf k} \over (2\pi )^n}
e^{i({\bf k}\cdot {\bf x}-k^2t)}\hat u_0({\bf k})+  \\
\sum\limits_{j=1}^n\int\limits_{{\cal R}^{n-1}}{d{\bf k}_j\over 
(2\pi )^{n-1}}e^{i{\bf k}_j\cdot {\bf x}_j}
\int_{\gamma}{dk_j\over 2\pi i}\{ e^{i(k_j|x_j|-k^2t)} 
[\hat w_{0j}({\bf k}_j,-k^2)+isign(x_j)k_j\hat v_{0j}({\bf k}_j,-k^2)]- \\
e^{i(k_j|x_j-L_j|-k^2t)} 
[\hat w_{Lj}({\bf k}_j,-k^2)+isign(x_j-L_j)k_j\hat v_{Lj}({\bf k}_j,-k^2)]\},
\ea
\eea
where $d{\bf k}_j=dk_1..\check{dk_j}..dk_n$ and  
$\gamma =(i\infty ,0)\cup (0,\infty )$.

In view of the distinguished parity properties of the Fourier transforms 
in (\ref{eq:FTuboxn}), in the following we shall make use of the 
parity operators: 
\beq\label{eq:parityproj}
\Delta_{\pm}=\prod\limits_{l=1}^n(1{\pm}\hat\sigma_l),~~~
\Delta^{(j)}_{\pm}=
\prod\limits_{l=1\atop l\ne j}^n(1{\pm}\hat\sigma_l),
\eeq
where $\hat\sigma_j$ is the involution $\hat\sigma_j:~k_j~\to~-k_j$.

Suppose we are interested in solving the Dirichelet problem; 
applying the parity operator $\Delta_-$ to (\ref{eq:FTuboxn}) 
eliminates all $\hat w_0$'s: 
\bea
\ba{c}
\Delta_-\hat u({\bf k},q)=-\left(
\Delta_-[\hat f({\bf k},q)+i\hat u_0({\bf k})]+
2i\sum\limits_{j=1}^nk_j\Delta^{(j)}_-\hat v_{0j}({\bf k}_j,q)+\right.  \\
\left.
2i\sum\limits_{j=1}^n[\sin (k_jL_j)\Delta^{(j)}_-\hat w_{Lj}({\bf k}_j,q)-
k_j\cos (k_jL_j)\Delta^{(j)}_-\hat v_{Lj}({\bf k}_j,q)]\right)
/(q+k^2-i0),~({\bf k},q)\in {\cal A}.
\ea
\eea
To eliminate also the $\hat {w_L}'s$, the values of $k_j$ must be  
restricted to the discrete set $k_j=h_j:={\pi m_j\over L_j},
~~m_j\in{\cal Z}$, so that the original domain $\cal A$ is finally restricted 
to:
\beq
({\bf k},q)\in {\cal A}'=\{({\bf h},q);~q\in\bar{\cal I}_q,~
{\bf h}=(h_1,..,h_n),~~
h_j={\pi m_j\over L_j},~~m_j\in{\cal Z},~~j=1,..,n\}.
\eeq
Therefore the EbR operator $\cal E$ of this example reads:
\beq
{\cal E}\cdot =\int_{{\cal R}^n}d{\bf k}\delta ({\bf k}-{\bf h})\Delta_-\cdot  
\eeq 
and its application to $\hat u$ leads to the wanted result:
\bea\label{eq:elimboxn}
\ba{c}
{\cal E}\hat u({\bf k},q)=(\Delta_-\hat u)({\bf h},q)=-\{
\Delta_-[\hat f({\bf h},q)+i\hat u_0({\bf h})]+  \\
2i\sum\limits_{j=1}^nh_j\Delta^{(j)}_-[\hat v_{0j}({\bf h}_j,q)-
(-)^{m_j}\hat v_{Lj}({\bf h}_j,q)]
\}/(q+h^2-i0),
\ea
\eea
where ${\bf h}_j=(h_1,..,\check{h}_j,..,h_n)$ and $h^2={\bf h}\cdot{\bf h}$.  
The transform ${\cal E}\hat u({\bf k},q)$ generated by the EbR procedure is 
the well-known multidimensional {\bf discrete} sine transform:
\bea\label{eq:dSFT}
\ba{c}
\Delta_-\hat u({\bf h},q)=
\int_{\cal D}dtd{\bf x}\tilde\varphi_{{\bf h},q}({\bf x},t)
u({\bf x},t),  \\
\tilde\varphi_{{\bf h},q}({\bf x},t):=
\Delta_-(e^{-i({\bf h}\cdot {\bf x}+qt)})=
(-2i)^ne^{-iqt}\prod\limits_{l=1}^n\sin (h_lx_l). 
\ea
\eea
For its inversion we use:
\beq\label{eq:eigenfdSFT}
\varphi_{{\bf h},q}({\bf x},t)=
{i^n\over L_1\cdot\cdot L_n}
{e^{iqt}\over 2\pi}\prod\limits_{l=1}^n\sin (h_lx_l),~~~~~~~~ 
\sum_{{\bf k},q}=
\int\limits_{\cal R}dq
\sum\limits_{m_1=1}^{\infty}\cdot\cdot\sum\limits_{m_n=1}^{\infty}~,
\eeq
so that (\ref{eq:U}) yields the function $\tilde U({\bf x},t)$, defined in 
the whole space time, which coincides with the solution 
$u({\bf x},t)$ of the IBV problem under scrutiny 
for $({\bf x},t)\in{\cal D}$: 
\bea\label{eq:UboxnD}
\ba{c}
\tilde U({\bf x},t)={i^n\over L_1\cdot\cdot L_n}
\sum\limits_{m_1=1}^{\infty}\cdot\cdot\sum\limits_{m_n=1}^{\infty}
\prod\limits_{l=1}^n\sin (h_lx_l)
\int\limits_{\cal R}{dq\over 2\pi}e^{iqt}
\Delta_-\left({\hat N({\bf h},q)\over {\cal L}(i{\bf h},iq)}\right),~~~
({\bf x},t)\in{\cal R}^{n+1}, \\
\tilde U({\bf x},t)=u({\bf x},t).~~~~({\bf x},t)\in{\cal D}.
\ea
\eea
Equation (\ref{eq:UboxnD}) implies, due to 
the symmetry properties of $\varphi_{{\bf h},q}({\bf x},t)$, that 
$\tilde U$ is the odd $(2L)$-periodic extension of the solution outside 
$\cal D$ and provides the discrete sine-Fourier representation of the solution:
\bea\label{eq:dSFrepr}
\ba{c}
u({\bf x},t)=-{i^n\over L_1\cdot\cdot L_n}\{
\sum\limits_{{\bf h},S}
\prod\limits_{l=1}^n\sin (h_lx_l)[
\int\limits_{\cal R}{dq\over 2\pi}e^{iqt}
{\Delta_-\hat f({\bf h},q)\over q+h^2-i0}-
e^{-ih^2t}\Delta_-\hat u_0({\bf h})]+  \\
\sum\limits_{j=1}^nL_j
\sum\limits_{{\bf h}_j,S}\prod\limits_{l\ne j}\sin (h_lx_l)
\int\limits_{\gamma}{dk_j\over \pi}
{ik_je^{-i(k^2_j+{\bf h}_j\cdot{\bf h}_j)t}\over \sin (k_jL_j)}
\Delta^{(j)}_-
[\sin k_j(L_j-x_j)\hat v_{0j}({\bf h}_j,-k^2_j-{\bf h}_j\cdot{\bf h}_j)+ \\
\sin (k_jx_j)\hat v_{Lj}({\bf h}_j,-k^2_j-{\bf h}_j\cdot{\bf h}_j)]
\},~~~~~~~~~~~~~~~~~~~~~~~~~({\bf x},t)\in
{\cal D},
\ea
\eea
where
\beq
\sum\limits_{{\bf h},S}=
\sum\limits_{m_1=1}^{\infty}\cdot\cdot\sum\limits_{m_n=1}^{\infty},~~~~~~~~~~~
\sum\limits_{{\bf h}_j,S}=
\sum\limits_{m_1=1}^{\infty}\cdot\cdot\left(
\check{\sum\limits_{m_j=1}^{\infty}}
\right)\cdot\cdot \sum\limits_{m_n=1}^{\infty}
\eeq
and the integral $\int_{\gamma}dk_j$ is regularized moving the singularities 
$k_j=(\pi m_j/L_j),~m_j\in{\cal Z}^+$ a bit off the first Quadrant. 

The Neumann problem can be treated similarly, leading to the discrete 
cosine transform:
\bea
\ba{c}
{\cal E}\cdot =\int_{{\cal R}^n}d{\bf k}\delta ({\bf k}-{\bf h})\Delta_+\cdot
  \\
\tilde\varphi_{{\bf h},q}({\bf x},t):=
\Delta_+(e^{-i({\bf h}\cdot {\bf x}+qt)})=
2^ne^{-iqt}\prod\limits_{l=1}^n\cos (h_lx_l),  \\
\varphi_{{\bf h},q}({\bf x},t)=
{1\over 2^nL_1\cdot\cdot L_n}
{e^{iqt}\over 2\pi}\prod\limits_{l=1}^n\cos (h_lx_l),~~~ 
\sum\limits_{{\bf k},q}=
\int\limits_{\cal R}dq
\sum\limits_{m_1=-\infty}^{\infty}\cdot\cdot
\sum\limits_{m_n=-\infty}^{\infty}
\ea
\eea
and to the following discrete cosine representation of the solution:
\bea\label{eq:dCFrepr}
\ba{c}
u({\bf x},t)=-{1\over 2^nL_1\cdot\cdot L_n}\{
\sum\limits_{{\bf h},C}
\prod\limits_{l=1}^n\cos (h_lx_l)[
\int\limits_{\cal R}{dq\over 2\pi}e^{iqt}
{\Delta_+\hat f({\bf h},q)\over q+h^2-i0}-
e^{-ih^2t}\Delta_+\hat u_0({\bf h})]-  \\
2\sum\limits_{j=1}^nL_j
\sum\limits_{{\bf h}_j,C}\prod\limits_{l\ne j}\cos (h_lx_l)
\int\limits_{\gamma}
{dk_j\over \pi}
{e^{-i(k^2_j+{\bf h}_j\cdot{\bf h}_j)t}\over \sin (k_jL_j)}
\Delta^{(j)}_+
[\cos k_j(L_j-x_j)\hat w_{0j}({\bf h}_j,-k^2_j-{\bf h}_j\cdot{\bf h}_j)- \\
\cos (k_jx_j)\hat w_{Lj}({\bf h}_j,-k^2_j-{\bf h}_j\cdot{\bf h}_j)]\},
~~~~~~~~({\bf x},t)\in{\cal D},
\ea
\eea
where
\beq
\sum\limits_{{\bf h},C}=
\sum\limits_{m_1=-\infty}^{\infty}\cdot\cdot
\sum\limits_{m_n=-\infty}^{\infty},~~~~~~~~~~~
\sum\limits_{{\bf h}_j,C}=
\sum\limits_{m_1=-\infty}^{\infty}\cdot\cdot\left(
\check{\sum\limits_{m_j=-\infty}^{\infty}}
\right)\cdot\cdot \sum\limits_{m_n=-\infty}^{\infty}
\eeq
and the integral $\int_{\gamma}dk_j$ is regularized as in (\ref{eq:dSFrepr}).

Using the convolution theorem, one immediately recovers from equations 
(\ref{eq:dSFrepr}) and (\ref{eq:dCFrepr}) the Green's integral representation 
(\ref{eq:Gintrepr}) of the solution, corresponding respectively to the 
following retarded Dirichelet and Neumann Green's functions:
\bea
\ba{c}
G_{RD}({\bf x},t;{\bf x}',t')={2^ni\over L_1\cdot\cdot L_n}H(t-t')
\sum\limits_{m_1=1}^{\infty}\cdot\cdot
\sum\limits_{m_n=1}^{\infty}e^{-ih^2(t-t')}
\prod\limits_{l=1}^n\sin (h_lx_l)\sin (h_lx'_l),     \\
G_{RN}({\bf x},t;{\bf x}',t')={-i\over  L_1\cdot\cdot L_n}H(t-t')
\sum\limits_{m_1\in{\cal Z}}\cdot\cdot
\sum\limits_{m_n\in{\cal Z}}e^{-ih^2(t-t')}
\prod\limits_{l=1}^n\cos (h_lx_l)\cos (h_lx'_l).
\ea
\eea

In the case of semicompact domains
the proper spectral transform generated by the EbR approach has, in 
general,  a continuous support. For instance, in the Dirichelet problem for 
the Schr\"odinger equation on the $n$-Quadrant
\beq\label{eq:nquadr}
V=\{{\bf x}\in{\cal R}^n:~x_j\ge 0,~j=1,..,n\},
\eeq
the elimination operator ${\cal E}=\Delta_-$, 
which restricts the definition domain to 
${\cal A}'={\cal R}^n\otimes \bar{\cal I}_q$, leads to the 
continuous sine Fourier transform:
\beq
\tilde\varphi_{{\bf k},q}({\bf x},t)=
(-2i)^ne^{-iqt}\prod\limits_{l=1}^n\sin (k_lx_l),~~
\varphi_{{\bf k},q}({\bf x},t)=
{e^{i({\bf k}\cdot{\bf x}+qt)}\over (2\pi )^{n+1}},~~ 
\sum_{{\bf k},q}=\int\limits_{\cal R}dq\int\limits_{{\cal R}^n}d{\bf k}~
\eeq
and to the continuous multidimensional sine-Fourier representation of 
the solution:
\bea\label{eq:SFrepr}
\ba{c}
u({\bf x},t)=
-\int_{{\cal R}^{n+1}}{dqd{\bf k} \over (2\pi )^{n+1}}
e^{i({\bf k}\cdot {\bf x}+qt)}
{\Delta_-\hat f({\bf k},q)\over q+k^2-i0}
+\int_{{\cal R}^n}{d{\bf k} \over (2\pi )^n}
e^{i({\bf k}\cdot {\bf x}-k^2t)}\Delta_-\hat u_0({\bf k})+  \\
\sum\limits_{j=1}^n\int\limits_{{\cal R}^{n-1}}{d{\bf k}_j\over 
(2\pi )^{n-1}}
\int_{\gamma}{dk_j\over \pi}e^{i({\bf k}\cdot {\bf x}-k^2t)} 
k_j\Delta^{(j)}_-\hat v_{0j}({\bf k}_j,-k^2),~~({\bf x},t)\in{\cal D}.
\ea
\eea

From the above illustrative examples it appears that 
the EbR procedure in Fourier space is very effective and, 
perhaps, simpler than the Green's function  approach, which is 
its counterpart in 
configuration space. The comparison between these two methods of 
elimination in examples in which the GF approach fails is  
postponed to a subsequent paper.  

\section{\bf The Elimination-by-Restriction Approach: 
\break The Nonlinear Case}

In this section we turn our attention to IBV problems associated with 
nonlinear evolution PDE's which are integrable by the inverse scattering 
(spectral) transform method. Because of the limited scope of this paper, we 
content ourselves with illustrating our method of solution by considering, as 
a prototype, the nonlinear Schr\"odinger (NLS) equation
\beq\label{eq:NLS}
iq_t+q_{xx}+c|q|^2q=0,~~q=q(x,t),
\eeq
where $c$ is an arbitrary real parameter, but this method applies as well 
to other $1+1$ dimensional soliton equations (f.i. to the Korteweg de Vries 
(KdV) equation). Moreover, we confine our treatment below to solutions 
of (\ref{eq:NLS}) in the first quadrant of the $(x,t)$ plane, namely on the 
semiline $0\le x\le \infty$ for $t\ge 0$. Besides the initial value  
$q(x,0)=q_0(x)$, the boundary value which uniquely specifies the solution is 
\beq\label{eq:BVf}
f(t)=a_1v(t)+a_2w(t),~~~~t\ge 0,
\eeq
where we have set
\beq\label{eq:BVvw}
v(t)=q(0,t),~~~~w(t)=q_x(0,t).
\eeq  
Here $a_1$ and $a_2$ are given real constants and, if $a_2=0$ ($a_1=0$) this 
is the Dirichelet (Neumann) BV problem. Thus the problem is that of 
constructing the solution $q(x,t)$ of (\ref{eq:NLS}) when the initial value 
$q_0(x)$ and the boundary value $f(t)$ are given functions in an appropriate 
functional space (we may assume that they are complex valued functions which 
rapidly decay as $x\to\infty$ and $t\to\infty$). 

The key-property of the NLS equation (\ref{eq:NLS}) is that it is the 
integrability condition for the following pair of $2\times 2$ matrix linear 
Ordinary Differential Equations (ODEs),
\beq\label{eq:Lax}
\Psi_x=(ik\sigma_3+Q)\Psi ,~~~~~\Psi_t=2ik^2[\sigma_3,\Psi ]+M\Psi
\eeq 
where $\sigma_3=diag~(1,-1)$ and 
\beq\label{eq:QM}
Q(x,t)= \left(
\begin{array}{cc}
0 & -c\bar q(x,t)  \\
q(x,t) & 0
\end{array}
\right),~~~~~~M(x,t,k)= 2kQ-i\sigma_3Q_x+iQ^2\sigma_3.
\eeq
The solution $\Psi (x,t,k)$ of these equations is defined by the asymptotic 
condition
\beq\label{eq:Jost}
\Psi (x,t,k)e^{-ikx\sigma_3}\to I,~~~~x\to\infty
\eeq
which uniquely defines the scattering matrix $S(k,t)$ in the standard way, 
namely as the boundary value
\beq\label{eq:S}
S(k,t)=\Psi (0,t,k).
\eeq

Well-known facts, which will be instrumental in the method below, are the 
following. The matrix solution $\Psi$ and, therefore (see (\ref{eq:S})), the 
scattering matrix $S$ have unit determinant,
\beq\label{eq:det1}
det~\Psi (x,t,k)=det~S(k,t)=1.
\eeq 
Moreover, the property
\beq\label{eq:symQ}
Q^{\dag}=-CQC^{-1},~~~~C:=\left(
\ba{cc}
1 & 0 \\
0 & c 
\ea
\right)
\eeq
of the matrix $Q$, see (\ref{eq:QM}), induces the corresponding property
\beq\label{eq:symPsi}
\Psi^{\dag}(x,t,k)=C\Psi^{-1}(x,t,\bar k)C^{-1},~~~
S^{\dag}(k,t)=CS^{-1}(\bar k,t)C^{-1},
\eeq
on the Jost solution $\Psi$ and on the scattering matrix $S$ 
(the superscript $^{\dag}$ indicates hermitian conjugation). As a 
consequence, it is convenient to parametrize the matrix $S$ by introducing the 
two functions $\alpha (k,t)$ and $\beta (k,t)$ according to the definition
\beq\label{eq:alphabeta}
S(k)=\left(
\ba{cc}
\alpha (k) & -c{\bar\beta (\bar k)} \\
\beta (k) & {\bar\alpha (\bar k)}
\ea
\right).
\eeq 
As for the $k$-dependence (here the complex spectral variable $k$ plays the 
same role as the Fourier variable in the linear case), the functions 
$\alpha (k,t)$ and $\beta (k,t)$ turn out to be analytic in the UHP ($Im~k>0$) 
and to have there the asymptotic behaviour
\beq\label{eq:asympt}
\alpha (k,t)=1+O(k^{-1}),~~~~\beta (k,t)=O(k^{-1}),
\eeq
for large $|k|$. Finally, we remind the reader that these analyticity 
properties of $\alpha (k,t)$ and $\beta (k,t)$ provide the way to solve the 
inverse problem, i.e. $S(k,t)\to Q(x,t)$, for any fixed $t\ge 0$; the basic 
equations of the inverse problem, which are not reported here, read as either 
Cauchy-type integral equations in the $k$ variable or, equivalently, as 
Marchenko-type integral equation in the $x$-variable.

Let us now look at the time evolution. Here the real crux of the spectral 
method appears in the evolution equation of the scattering matrix, see 
(\ref{eq:Lax}) and (\ref{eq:S}),
\beq\label{eq:evolS}
S_t=2ik^2[\sigma_3,S]+Z(k,t)S,
\eeq     
since the matrix $Z(k,t)$ has a separate dependence on both the boundary data 
$v(t)$ and $w(t)$ (see (\ref{eq:BVvw})) according to the following expressions
\bea\label{eq:Z}
\ba{c}
Z(k,t)=2kV(t)-i\sigma_3W(t)+iV^2(t)\sigma_3,  \\
V(t)=Q(0,t),~~~~W(t)=Q_x(0,t).
\ea
\eea
As a consequence, the evolution equation (\ref{eq:evolS}) cannot be 
immediately 
integrated to yield the scattering matrix whose knowledge is essential to 
reconstruct $Q(x,t)$ via the solution of the inverse problem. Since 
only the boundary datum (\ref{eq:BVf}) is given, in analogy with the 
elimination strategy of the linear 
case, we introduce at this point the novel matrix
\beq\label{eq:Stilde}
\tilde S(k,t)=A^{-1}(k)S^{-1}(-k,t)A(k)S(k,t),~~~A(k):=a_1I+2ika_2\sigma_3,
\eeq
because of its two important properties. First, its determinant is unit 
(see (\ref{eq:det1})) and its asymptotic value as $|k|\to\infty$ is the unit 
matrix:
\beq\label{eq:detStilde}
det~\tilde S(k,t)=1,~~~\tilde S(k,t)=I+O(k^{-1});
\eeq
second, it satisfies an evolution equation which contains only the given 
boundary value (\ref{eq:BVf}), namely
\bea\label{eq:evolStilde}
\ba{c}
\tilde S_t=2ik^2[\sigma_3,\tilde S]+4kA^{-1}(k)S^{-1}(-k,t)F(t)S(k,t),  \\
F(t)=a_1V(t)+a_2W(t)=\left(
\ba{cc}
0 & -c\bar f(t) \\
f(t) & 0
\ea
\right).
\ea
\eea
Though this is an important step, it does not yield the solution of our 
problem since the unknown scattering matrix $S(k,t)$ still appears in the 
evolution (\ref{eq:evolStilde}a). Thus one has to find the way to relate 
$S(k,t)$ and  
$\tilde S(k,t)$ to each other. The relation $S(k,t)\to\tilde S(k,t)$ is of 
course trivial as it is 
given by the definition (\ref{eq:Stilde}) itself. This relation yields the 
initial 
value $\tilde S(k,0)$ for the integration of the evolution equation 
(\ref{eq:evolStilde}), i.e. $Q(x,0)\to \Psi (x,0,k)\to S(k,0)=\Psi (0,0,k)\to 
\tilde S(k,0)$. As for the inverse relation, $\tilde S(k,t)\to S(k,t)$, one 
has instead to set up a RH problem, which finally leads to a Cauchy-type 
integral equation. Starting with rewriting (\ref{eq:Stilde}) in the form 
$A(k)S(k,t)=S(-k,t)A(k)\tilde S(k,t)$, and noting that the first column of   
$S(k,t)$ in (\ref{eq:alphabeta}) is analytic in the UHP with the asymptotic 
behaviour 
(\ref{eq:asympt}), one can write down two coupled integral equations for 
$\alpha (k,t)$ and $\beta (k,t)$ in terms of $\tilde S(k,t)$ by going through 
the standard RH problem technique. By assuming, just for the sake of 
simplicity, that no poles occur in the UHP, and by substituting 
$\tilde S(k,t)$ with its expression obtained 
by formally integrating the 
evolution equation (\ref{eq:evolStilde}), one finally ends up with the 
two coupled nonlinear integral equations
\bea\label{eq:RH}
\ba{c}
\alpha (k,t)=1+{c\over 2\pi i}\int\limits_{-\infty}^{\infty}
{dk'\over k'-(k+i0)}e^{-4i{k'}^2t}h(k',t)\bar\beta (k',t),  \\
\beta (k,t)=-{1\over 2\pi i}\int\limits_{-\infty}^{\infty}
{dk'\over k'-(k+i0)}e^{-4i{k'}^2t}h(k',t)\bar\alpha (k',t), 
\ea
\eea
whose nonlinearity is due to the fact that the function $h(k,t)$ depends 
itself on the unknowns $\alpha (k,t)$ and $\beta (k,t)$ through the integral 
(in t) relation
\bea\label{eq:defh}
\ba{c}
h(k,t)=
\{a(k)\alpha_0(k)\beta_0(-k)-a(-k)\alpha_0(-k)\beta_0(k)-  \\
4k\int\limits_0^tdt'e^{4ik^2t'}[f(t')\alpha (k,t')\alpha (-k,t')+
c\bar f(t')\beta (k,t')\beta (-k,t')]
\} /  \\
\{a(-k)\alpha_0(-k)\bar\alpha_0(k)+ca(k)\beta_0(-k)\bar\beta_0(k)-  \\
-4kc\int\limits_0^tdt'[f(t')\alpha (-k,t')\bar\beta (k,t')-
\bar f(t')\bar\alpha (k,t')\beta (-k,t')]\},
\ea
\eea  
where $a(k)=a_1+2ika_2$ and $\alpha_0$ and $\beta_0$ are, respectively, the 
known initial values $\alpha (k,0)$ and $\beta (k,0)$.

The discussion of these, admittedly complicate, equations, and of their 
implications in various directions is beyond the scope of this paper, and 
will be reported elsewhere. Here we merely note that this formulation 
naturally single out the so-called linearizable IBV problems, these being 
those for which the boundary value $f(t)$, see (\ref{eq:BVf}), vanishes: 
$f(t)=0$. Indeed, in this case, the kernel function $h(k,t)$ (\ref{eq:defh}) 
does not depend on the unknowns  $\alpha (k,t)$ and $\beta (k,t)$, and the 
equations (\ref{eq:RH}) become linear. As a side remark, we also note that 
setting $c=0$ eliminates the nonlinearity in all the formulae given above, 
so that (\ref{eq:NLS}) becomes the linear Schr\"odinger equation and our 
equations (\ref{eq:RH}), (\ref{eq:defh}) yield $\alpha (k,t)=1$ while 
$\beta (k,t)$ coincides 
with the usual explicit expression of the Fourier transform of the solution 
$q(x,t)$.

Finally, we deem of interest to report here also an approach to the IBV 
problem for the NLS equation (\ref{eq:NLS}) which is different from the 
one given above, and yet equivalent as it eventually leads to the same 
equations (\ref{eq:RH}) and (\ref{eq:defh}). The main feature of this 
approach is that the IBV problem is reformulated on the whole line, i.e. for 
$x\in (-\infty ,\infty )$, and that the matrix $\tilde S(k,t)$, as defined 
by (\ref{eq:Stilde}), acquires now a spectral meaning within the standard 
direct and inverse problem associated with the Lax equation, say the first 
ODE in (\ref{eq:Lax}). The price one pays to arrive at this more familiar 
formulation is that the nonlinear PDE one has to solve now is the NLS 
equation with a inhomogeneous source term rather than the NLS equation 
(\ref{eq:NLS}). This approach is briefly sketched here with two limitations 
which are merely dictated by the sake of simplicity; namely we confine our 
treatment to the Dirichelet and Neumann BV problems and, second, we assume 
that the spectral data at any time $t\ge 0$ have no discrete spectrum 
component.

The starting observation is that, if $q(x,t)$ is a solution of the NLS 
equation (\ref{eq:NLS}) for $x\in (0,\infty )$ and $t\ge 0$, then the function
\beq\label{eq:qtilde}
\tilde q(x,t)=q(x,t)H(x)-\eta q(-x,t)H(-x),~~~\eta =\pm 1,
\eeq  
as defined for any real value of $x$, satisfies the PDE
\beq\label{eq:forcedNLS}
i\tilde q_t+\tilde q_{xx}+2c|\tilde q|^2\tilde q=(1+\eta )v(t)\delta'(x)+
(1-\eta )w(t)\delta (x). 
\eeq
where $\delta (x)$ is the Dirac delta distribution,  
$\delta'(x)$ is its derivative and 
$v,~w$ are defined in (\ref{eq:BVvw}). Obviously, $\eta =1$ ($\eta =-1$) 
is the appropriate choice when one deals with the Dirichelet 
(Neumann) IBV problem.

As implied by the spectral method based on the Lax equations,  
it is convenient to rewrite (\ref{eq:qtilde}) and (\ref{eq:forcedNLS}) in 
matrix form by introducing the $2\times 2$ off-diagonal matrix (see 
(\ref{eq:QM}))
\beq\label{eq:Qtilde}
\tilde Q(x,t)=Q(x,t)H(x)-\eta Q(-x,t)H(-x),~~~\eta =\pm 1,
\eeq  
and the PDE
\beq\label{eq:forcedmatrixNLS}
i\tilde Q_t-\sigma_3(\tilde Q_{xx}-2\tilde Q^3)=\Sigma (x,t) 
\eeq
which is, of course, equivalent to (\ref{eq:forcedNLS}) if the source term 
is (see (\ref{eq:Z}b))
\beq\label{eq:defSigma}
\Sigma (x,t)=-\sigma_3[(1+\eta )V(t)\delta'(x)+(1-\eta )W(t)\delta (x)].
\eeq
The spectral approach to the equation (\ref{eq:forcedmatrixNLS}) is based 
on the spectral equation
\beq\label{eq:Laxtilde}
\tilde\Psi_x=(ik\sigma_3+\tilde Q(x,t))\tilde\Psi ,
~~~~~\tilde\Psi=\tilde\Psi (x,t,k),
\eeq
and it is standard. The Jost solution $\tilde\Psi$ is defined by the 
asymptotic condition (\ref{eq:Jost}), $\tilde\Psi exp(-ikx\sigma_3)
\to I,~x\to\infty$, which readly provides its expression in terms of the 
solution $\Psi (x,t,k)$ introduced above on the semiline,
\beq\label{eq:Psitilde}
\tilde\Psi (x,t,k)=\Psi (x,t,k)H(x)+E\Psi (-x,t,-k)E\tilde S(k,t)H(-x),
\eeq 
where $E=diag~(1,\eta )$ and 
\beq\label{eq:SStilde}
\tilde S(k,t)=ES^{-1}(-k,t)ES(k,t),
\eeq
is precisely the scattering matrix which is defined in the usual way, namely
\beq\label{eq:Josttilde}
\tilde\Psi (x,t,k)\to e^{ikx\sigma_3}\tilde S(k,t),~~x\to -\infty .
\eeq
At this point we note that this scattering matrix $\tilde S(k,t)$ coincides 
with the matrix (\ref{eq:Stilde}) with $a_2=0$ in the Dirichelet case 
($\eta =1$) 
and with $a_1=0$ in the Neumann case ($\eta =-1$).

It is common expedient now to introduce also the other Jost solution of 
(\ref{eq:Laxtilde}),
\beq\label{eq:Phitilde}
\tilde\Phi (x,t,k)=\tilde\Psi (x,t,k){\tilde S}^{-1}(k,t),
\eeq   
and to take into account the identity
\beq\label{eq:tildeid}
\tilde S_t+2ik^2[\tilde S,\sigma_3]=i\int\limits_{-\infty}^{\infty}dx
{\tilde\Phi}^{-1}(x,t,k)[i\tilde Q_t-\sigma_3(\tilde Q_{xx}-2{\tilde Q}^3)]
\tilde\Psi (x,t,k), 
\eeq
which, together with the inhomogeneous PDE (\ref{eq:forcedmatrixNLS}), 
entails the evolution equation for the scattering matrix,
\beq\label{eq:evolStildebis}
\tilde S_t=2ik^2[\tilde S,\sigma_3]+i\int\limits_{-\infty}^{\infty}dx
{\tilde\Phi}^{-1}(x,t,k)\Sigma (x,t)
\tilde\Psi (x,t,k).
\eeq
It is now easy to show that inserting in the integral in the RHS of this 
equation the expressions (\ref{eq:Phitilde}), (\ref{eq:Psitilde}) and 
(\ref{eq:defSigma}) yields 
precisely the evolution equation (\ref{eq:evolStilde}) for the Dirichelet 
and Neumann IBV problems, say
\beq\label{eq:evolStilde2}
\tilde S_t=2ik^2[\sigma_3,\tilde S]+2k(1+\eta )S^{-1}(-k,t)V(t)S(k,t)-
i(1-\eta )\sigma_3S^{-1}(-k,t)W(t)S(k,t).
\eeq 
We end this paper remarking that a good side of the present approach, 
which we will refer to as the 
``source-method'', is that one may take advantage of the more traditional 
inverse scattering (spectral) technique on the whole line. In particular, 
to investigate the large time behaviour of the solution $q(x,t)$ of 
the IBV problem since the asymptotic expression, if the boundary value 
rapidly vanishes as $t\to\infty$, are readly at hand in the usual spectral 
theory on the whole line.

\section{Literature}
 
The classical idea of eliminating unknown boundary values 
from the representation of the solution is the essence of the Green's  
function approach \cite{MF}, which makes essential use of the image method 
\cite{TS} to construct the proper Green's function which eliminates the 
unknown boundary values. The Elimination-by-Restriction approach 
introduced here  
is a natural and effective implementation of the elimination strategy 
in Fourier space. To the best of our knowledge this method for 
linear PDEs has never been presented before.

An alternative method, which we term the {\it Analyticity} approach, is also 
possible, and it is motivated by Fokas discovery 
of the global relation and of its use to solve IBV problems \cite{Fokas1}, 
\cite{Fokas2}, \cite{Fokas3}, \cite{FP}, \cite{Pelloni}. Our contribution 
to this method consists in  using 
systematically the analyticity properties of all the Fourier transforms 
involved in (\ref{eq:FTu}), to derive a set of analyticity constraints 
which allow one to express unknown boundary values in terms of known ones 
and, in general, to study the unique solvability of IBV problems (see 
\cite{DMS2}).

Different approaches to deal with the problem of unknown boundary data in 
the study of IBV problems for soliton equations 
have been developed  during the last few decades. In \cite{Fokas4}, Fokas 
introduced a nonlinear analogue of the sine transform.  
In \cite{Sabat}, Sabatier constructed an ``elbow 
scattering'' in the $(x,t)$- plane to deal with the semiline problem for KdV, 
leading to a 
Gel'fand - Levitan - Marchenko formulation. In \cite{Fokas1,Fokas2} a 
different approach, based on a simultaneous x-t spectral transform, 
has been introduced  by Fokas and rigorously developed in \cite{FIS,Fokas5},  
to solve IBV problems for soliton equations on 
the semiline. It allows one for a rigorous asymptotics \cite{FI} and captures 
in a natural way the known cases \cite{linbv} of linearizable boundary 
value problems. In \cite{DMS1} we have introduced two alternative 
approaches to the study of IBV problems for soliton equations on the 
segment and on the semiline. In the first method we expressed the unknown 
boundary values in terms of elements of the scattering matrix $S(k,t)$, 
thus obtaining a nonlinear integro-differential evolution equation for $S$. 
In the second method, which can be viewed as the nonlinear analogue of the EbR 
approach developed in \S 2, we constructed the nonlinear evolution equation 
(\ref{eq:evolStilde}) for $S$, which does not contain unknown boundary 
values and captures in a natural way the case of linearizable IBV problems.  

In some nongeneric cases of soliton equations corresponding to singular 
dispersion relations, like the stimulated 
Raman scattering (SRS) equations and the sine Gordon (SG)   
equation in light-cone coordinates, the evolution equation of the scattering 
matrix does not contain unknown boundary data.   
The SG equation on the semiline has been treated using the 
x-t spectral transform \cite{Fokas1}; the SRS and the SG equations on 
the semiline have also been treated using a more traditional x- 
transform method respectively in \cite{LM} by Leon and Mikhailov 
and in \cite{LS} by Leon and Spire; the x- 
spectral data used in this last approach satisfy a nonlinear evolution 
equation of Riccati type.   

Apart from the simultaneous $x-t$ transform, all the above approaches 
are based on the traditional IST \cite{libri}.   
The spectral formalism for studying forced soliton equations has been 
developed by several authors, expecially in connection to the 
theory of perturbations (see, for instance, \cite{Newell}).  

The results presented here in the nonlinear context 
are: i) the construction, via the RH 
technique, of the closed form integral equation (\ref{eq:RH}), (\ref{eq:defh}) 
satisfied by the 
elements of the scattering matrix. ii) The formulation of the Dirichelet 
and Neumann IBV 
problems on the semiline for soliton equations as forced initial value 
problems on the whole line. The equivalence between the semiline and 
whole-line problems has been already used in \cite{Fokas4}, although the 
relevant 
equations for the spectral data given there differ from those presented here.

\vskip 5pt 
\noindent
{\bf Acknowledgments}

\vskip 1pt
\noindent
The present work has been carried out during several visits and meetings. 
We gratefully acknowledge the financial contributions provided by the 
RFBR Grant 01-01-00929, the INTAS Grant 99-1782 and by the 
following Institutions: the University of Rome ``La Sapienza'' (Italy), 
the Istituto Nazionale di Fisica Nucleare (Sezione di Roma), the Landau 
Institute for 
Theoretical Physics, Moscow (Russia) and the Isaac Newton Institute,  
Cambridge (UK), within the programme ``Integrable Systems''.


\begin{thebibliography}{99}

\bibitem{MF} P. M. Morse and H. Feshbach, {\it Methods of Theoretical 
Physics}, Mc Graw-Hill, New York (1953). 

\bibitem{TS} R. Terras and R. Swanson, J.Math.
Phys. {\bf 21}, 2140-2153 (1980).


\bibitem{Fokas1} A. S. Fokas, Proc. Roy. Soc. Lond. A, {\bf 53}, 1411 (1997).

\bibitem{Fokas2} A. S. Fokas, J. Math. Phys. {\bf 41}, 4188 (2000).

\bibitem{Fokas3} A. S. Fokas, Proc. Roy. Soc. Lond. A, {\bf 457}, 371 (2001).

\bibitem{FP} A. S. Fokas and B. Pelloni, Math. Proc. Camb. Phil. Soc., 
{\bf 131}, 521 (2001).

\bibitem{Pelloni} B. Pelloni, ``On the well-posedness of boundary value 
problems for integrable evolution equations on a finite interval'', in the  
Proceedings of NEEDS 2001; editors A. Mikhailov and P. M. Santini. To be 
published in Theor. Math. Phys., 2002. 

\bibitem{DMS2} A. Degasperis, S. V. Manakov and P. M. Santini, ``Initial-
Boundary Value Problems: Spectral Methods of Solution'', Preprint 2002.

\bibitem{Fokas4} A. S. Fokas, ``Inverse scattering transform on the half line 
- the Nonlinear analogue of the sine transform'', in {\it Inverse Problems: 
An Interdisciplinary Study}, edited by P. C. Sabatier, Academic Press, 
London, 1987.

\bibitem{Sabat} P. C. Sabatier, J. Math. Phys. {\bf 41}, 414 (2000).


\bibitem{FIS} A. S. Fokas, A. R. Its and L. Y. Sung, preprint 2001.

\bibitem{Fokas5} A. S. Fokas, ``Integrable Nonlinear Evolution Equations on 
the Half-Line'', preprint (October 2001). 

\bibitem{FI} A. S. Fokas and A. R. Its, Phys. Rev. Lett. {\bf 68}, 3117 (1992).
\bibitem{linbv} M. J. Ablowitz and H. Segur, J. Math. Phys. {\bf 16},  
1054 (1975).

\bibitem{DMS1} A. Degasperis, S. V. Manakov and P. M. Santini, JETP Letters 
{\bf 74}, 541 (2001).

\bibitem{LM} J. Leon and A. V. Mikhailov, Phys. Lett. A {\bf 253}, 33 (1999).

\bibitem{LS} J. Leon and A. Spire, Preprint nlin.PS/0105066. 

\bibitem{libri} M.J.Ablowitz and H.Segur, {\it Solitons and the inverse 
scattering transform} SIAM, Philadelphia, 1981; V.E.Zakharov, 
S.V.Manakov,S.P.Novikov and L.P.Pitaevsky, {\it Theory of solitons}, 
Nauka, 1980 [Consultants Bureau (Plenum Publ.), New York, 1984].

\bibitem{Newell} A. C. Newell, ``The Inverse Scattering Transform'', in 
{\it Solitons}, edited by R. K. Bullough and P. J. Caudrey, Springer-Verlag, 
Berlin, 1980.


\end{thebibliography}
\end{document}